\documentclass[aps,twocolumn,pra,showpacs]{revtex4}
\usepackage[dvips]{epsfig}

\date{\today}

\begin{document}

\title{On the Possibility of the Strong Field QED Investigation at LHC}

\author{V.~G.~Baryshevsky \footnote {bar@minsk.inp.by} and V.~V.~Tikhomirov \footnote {vvtikh@mail.ru}}
\address
{Research Institute for Nuclear Problems, Belarussian State
University, Bobruiskaya 11, 220050 Minsk, Belarus}

\draft
\begin{abstract}
The interaction of high energy particles with atomic axes and
planes allows to observe in crystal various effects predicted by
the quantum electrodynamics of phenomena in strong electromagnetic
field. In particular, the effect of electron-positron pair
production by gamma-quanta in a semi-uniform field was observed
for the first time in eightieth in CERN in the field of germanium
crystal axes. The high energy of LHC drastically widens the
possibilities of strong field QCD effect investigation in crystals
allowing to observe vacuum dichroism and birefringence, electron
radiative self-polarization and polarized electron-positron pair
production by gamma-quanta, positron (electron) anomalous magnetic
moment modification and electron spin rotation in crystal field
harmonics. The effect of vacuum birefringence induced by strong
electric field is considered in detail.
\end{abstract}
\pacs{PACS: 13.40.-f, 78.20.Fm, 23.20.Ra, 41.60.-m, 61.85.+p}
 \maketitle

%*****************   The Body of the Article:   *************************

\section{Introduction}
The effects induced or strongly modified by intense
electromagnetic fields greatly widen our understanding of the
nature of electromagnetic field, electromagnetic interaction and
particle structure. These effects play important role in
dense cosmic objects as well as in cosmic ray generation.

For a long time it seemed impossible to study the effects of QED
in strong fields in earth conditions. Even the unique experiments
of Kapitsa and Sakharov made it possible to generate only the
fields millions times lower than the critical fields
\begin {equation}%1
H_0 = m^2c^3/|e|\hbar=4.4\times 10^{13} G,
\end{equation}
\begin {equation}%2
E_0 = m^2c^3/|e|\hbar=1.32\times 10^{16} V/m,
\end{equation}
where $m$ is electron mass, typical for the atmospheres of neutron
stars and magnetars in which such effects of strong field QED as
electron-positron pair production by gamma-quanta, gamma-quanta
birefringence and splitting, electron magnetic moment modification
and electron spin radiative self-polarization play important role.
However in seventieth and eighties the possibility of observation
of the effects of QED in strong uniform crystal fields was
realized \cite{bar1,kim,sor,bai,ugg,art}.

A wide applicability of the uniform field approximation for
description of radiation processes in the fields of crystal planes
and axes is explained by the fact that a forming gamma-quantum or
electron-positron pair moves inside a formation region which, on
the one hand, has a length several orders exceeding the
interatomic distance and, on the other, a  transverse dimension
which can be as small as the Compton length. The great length of
formation region provides a smearing out of atomic potentials to a
continuum potential remaining invariant under the translations
along crystal plane or axis. In its turn, the small transverse
dimension provides nearly constant value of the field acting
during formation process on particles moving at sufficiently small
angles in respect to the crystal plane or axis.

The applicability of the uniform field approximation to
description of process in oriented crystals allowed us to predict
\cite{bar2,bar3} the pair production process in strong
semi-uniform continuum crystal field as well as intimately
connected to it strong crystal optical anisotropy in hard
gamma-region, directly connected with the anisotropy of strong
uniform field in vacuum. While the pair production process itself
was observed and studied experimentally in CERN in middle
eightieth in the field of $Ge$ and $W$ crystal {\it axes}, the
energies $150-200 GeV$ of tertiary CERN gamma beams were that time
quite far from that needed to observe the crystal optical
anisotropy in hard gamma-region, namely, the effects of dichroism
and birefringence as well as many important effects connected with
electron spin, all of which manifest themselves in the fields of
crystal {\it planes} considerably less intensive then that of
crystal axes. The aim of this paper is to demonstrate that the LHC
energy provides really optimal conditions to observe the effects
of strong crystal optical anisotropy in hard gamma-region
\cite{bar2,bar1} directly connected with the optical anisotropy of
strong uniform field \cite{rit,erb,heis,schw,qed2}.

\section{Typical fields and energies}

Let us first demonstrate that the fields of crystal axes and
planes are really large. Remind  \cite{lin} that nuclei positions
in crystals are normally scattered by their equilibria with a mean
square displacement $u \simeq 0.05 - 0.1 \AA$. The scattering of
atomic nuclei positions suppresses the continuum potential growth
at distances $\Delta \rho \leq u$ from a plane formed by
equilibrium atomic positions limiting, thus, the maximum electric
field of a crystal axis by the value
\begin {equation}%3
E_{max}^{ax} \simeq 2Ze/u d_{ax} \simeq  10^{10} Z (V/cm) \leq
10^{12}V/cm,
\end{equation}
where $Z$ is the monocrystal atomic number and $d_{ax}$ in the
interatomic distance on the axis. The estimate (3) was obtained by
averaging a nucleus field over a strait line passing at distance
$u$. A very similar procedure of the averaging of a nucleus field
over a plane gives an order lower estimate
\begin {equation}%4
E_{max}^{pl} \simeq \pi Ze n_0 d_{pl}  \leq 10^{11}V/cm
\end{equation}
for the maximum value of the longitudinally averaged field of
atomic crystal plane, where $n_0$ and $d_{pl}$ are crystal atomic
number density and inter-plane distance, respectively. The results
of evaluation of maximum averaged fields of the crystals axes and
planes most widely used in channelling studies are given in the
Table. The gamma-quantum energies which are enclosed in brackets
correspond to deeply cooled crystals while all the others have
been obtained for the normal temperature. In their ability to
deflect relativistic particles the fields (3) and (4) are
equivalent, respectively, to magnetic fields of strengths $0.2-3
GG$ and $20-200 MG$, far exceeding that produced by cumulative
magnetic generators for microsecond durations only \cite{knop}.
\vspace{5mm}

Table
\\
\begin{tabular}{|c|c|c|c|}
\hline
   Crystal & (plane) & $E_{max}$, & $\omega_{0}$, \\
   & $<$Axis$>$ &GeV/cm & TeV \\
   \hline
  Diagonal & $($110$)$ & 7.7 & 1.78 \\
     & $<$110$>$ & 75 & 0.20 \\
  \hline
  Si & $($110$)$ & 5.7 & 2.39 (1.7) \\
     & $<$110$>$ & 46 & 0.29 \\
  \hline
  Ge & $($110$)$ & 9.9 & 1.37 (0.9) \\
    & $<$110$>$ & 78 & 0.174 \\
  \hline
  W & $($110$)$ & 43 & 0.316 \\
    & $<$111$>$ & 500 & 0.027 \\ \hline
\end{tabular}
\vspace{5mm}

Though the fields (3) and (4) are quite large, they are still far
below the field (2) capable to produce electron-positron pairs in
vacuum. Fortunately, such a field can be reached in the reference
frame of ultra relativistic particles in which the transverse
averaged crystal field is "multiplied" by the Lorenz factor equal
to the particle energy measured in the units of its rest energy.
Considering the pair production process which is intimately
connected with dichroism and birefringence phenomena, it will be
natural to use the mass $2 mc^2$ as a "rest mass" of a forming
electron-positron pair leading to the relativistically magnified
field strength $E_{max} \omega/2 m$ (below we use the system of
units in which $\hbar = c =1$) in its reference frame. This
strength will reach the critical value (2) at the gamma-quantum
energy
\begin {equation}%5
\omega_0 = 2m \frac{E_0}{E_{max}},
\end{equation}
which can be considered as a "threshold" energy of pair production
in crystals. Some values of this energy given in the Table explain
why the investigations of strong field QED effects in mid
eightieth were tightly restricted by the axial case, which allowed
to observe the electron-positron pair production in a semi-uniform
electric field, however has not made possible to observe the
effects of uniform field optical anisotropy as well as the number
of effects connected with electron spin. The Table also shows that
observation of these effects will become possible at the LHC
gamma- and electron beams the energies of which will definitely
exceed $1 TeV$.

\section{Crystal dichroism}

From the many strong field QED effects in the uniform field the
birefringence effect is naturally distinguished by its fundamental
nature, simplicity of theory and by the limited energy of its
optimal manifestation. This energy limit arises because of rapid
growth of gamma-quantum absorption at energies $\omega > \omega_0$
due to the electron-positron pair production, accompanied by the
strong dichroism. Both the dichroism and birefringence of uniform
field manifest themselves much better in the field of crystal
planes because the letter is parallel to the same line, the normal
to crystal planes, in all the crystal volume.

\begin{figure}[h!]% Fig.1.
\leavevmode \centering
\includegraphics[width=3in, height=2in, angle=0]{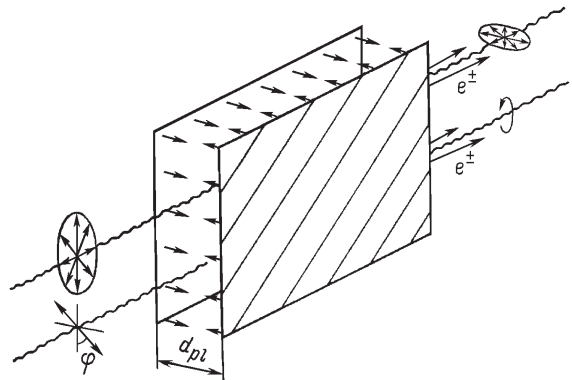}
\vspace*{-0.4cm}\caption{Averaged field of crystal planes allowing
to observe dichroism and birefringence effects inherent to strong
electromagnetic field in vacuum.} \label{nurfum}
\end{figure}

Various important aspects of theory of electron-positron pair
production in crystals and accompanying effects of dichroism and
birefringence were considered in
\cite{bar1,bar2,bar3,bar4,bar5,bar6,tikh1,tikh2} (see also
\cite{fro} and \cite{bai}, \S 15). First of all, it was shown
there that the general QED formulas can be considerably simplified
in the case of gamma-quantum incidence nearly parallel to crystal
planes when they reduce to the formulas of the theory of
electron-positron pair production by gamma-quanta in a uniform
electromagnetic field \cite{rit,erb,bai} averaged over the crystal
volume. Let us take the probability
\begin{eqnarray}%6
\nonumber W_{x,y}(\kappa)=\frac{\alpha m^2}{\sqrt{\pi}\omega ^2} \\
 \times\int _0^{\omega}d\varepsilon _{+}\biggl[\int
_{\xi}^{\infty}\Phi (y)dy+\biggl( 2\pm 1-\frac{\omega
^2}{\varepsilon _{+}\varepsilon
_{-}}\biggr)\frac{{\Phi}'(\xi)}{\xi}\biggr],
\end{eqnarray}
where
\begin {equation}%7
\kappa=\frac{E \omega}{E_0 m},\qquad \xi=\biggl(\frac{E_0
m\omega}{E\varepsilon _{+}\varepsilon _{-}}\biggr)^{2/3},
\end{equation}
$\varepsilon _{-}$ and $\varepsilon _{+}$ are the electron and
positron energies, respectively.
\begin {equation}%8
\Phi(\xi)=\frac{1}{\sqrt\pi}\int _0^{\infty}\cos\biggl(\xi
t+\frac{t^3}{3}\biggr)dt
\end{equation}
is Airy function and the transverse axes $x$ and $y$ are,
respectively, parallel and normal to the electric field strength.
According to \cite{bar2,bar3} the probabilities of pair production
by gamma-quanta propagating parallel to crystal planes and
polarized parallel and normal to them can be evaluated by the
simple formula
\begin {equation}%9
W^{coh}_{\parallel ,\perp}=\int
^{d_{pl}}_0W_{y,x}({\kappa}(E(x)))\frac{dx}{d_{pl}}
\end{equation}
in which the parameter $\kappa$ in arbitrary point $x$ should be
calculated substituting the averaged electric field strength
$E(x)$ of crystal planes in that point to eq. (7).

A direct averaging (9) of the local pair production probabilities
(6) over an inter-plane distance $d_{pl}$ drastically simplifies
the theoretical description of pair production and related crystal
optical anisotropy in hard gamma region. The limits of
applicability of this simplifying procedure were also
investigated. It was found \cite{bar6} that at very small
incidence angles $\theta \ll 1\mu rad$ the local gamma-quantum
flux attenuation near nuclear planes and axes violates an
applicability of the averaging procedure (9) and leads to a new
orientational effect which also can be observed at LHC. At large
incidence angles $\theta \geq 100\mu rad$ both the field
non-uniformity \cite{bar4} and interference of amplitudes of
radiation and pair production processes in the fields of different
planes (and axes) become important. An efficient numerical method
based on fast Fourier transform was developed in \cite{tikh2} to
evaluate the radiation and pair production characteristics in this
case. Most important for future studies is the conclusion that the
averaging procedure (9) (see also eq. (15) and (17) below) remans
valid in all the incidence angle region $1\mu rad \leq \theta \leq
100\mu rad$ representing the main practical interest for the
future investigations at LHC energies.

Since the Airy function (8) decreases like $exp{(-2 \xi^{3/2}/3)}$
at $\xi > 1$, the pair production is suppressed in all the crystal
at any distribution of gamma-quantum energy between the electron
and positron if $\xi > (3/2)^{2/3}$ at $\varepsilon _{+} =
\varepsilon _{-} = \omega /2$ and $E = E_{max}$. A "strict"
estimate of the "threshold" gamma-quantum energy following from
these conditions differs from the estimate (5) only by a
multiplier $8/9$.

Besides the necessity to overage the local probabilities (6) over
the crystal volume (in fact, over the inter-plane distance - see
eq. (9)), there exists another complicating feature of the pair
production process in crystals which does not accompany the same
process in a uniform electromagnetic field in vacuum. In fact, the
probabilities (9) describe only the coherent pair production
process, taking place in the averaged field of crystal planes. In
addition to it, the incoherent pair production process on separate
nuclei also manifests itself in crystals. This process reminds the
Bethe-Heitler pair production process, however is considerably
modified by the influence of the averaged crystal field on the
pair formation process initiated by a nucleus. The investigation
of the radiation and pair production processes in the uniform
field in the presence of Coulomb scatterers allowed us to combine
\cite{bar5} their theories in uniform electromagnetic field and in
amorphous medium as well as to obtain the expressions
\cite{bar1,tikh1}
\begin{eqnarray}%10
\nonumber W_{\parallel (\perp)}^{inc}=\frac{1}{15L_{rad}\omega ^3} \\
\nonumber \times\int
_0^{d_{pl}}\frac{n(x)}{n_0}\frac{dx}{d_{pl}}\int
_0^{\omega}d\varepsilon _{+}\biggl[1-\theta
(1-\xi)\frac{ln\xi}{2ln(183Z^{-1/3})}\biggr] \\
\nonumber \times \biggl[ \omega ^2[(1\pm 1)\xi ^4\Upsilon\mp
6\xi\Upsilon
-(3\pm 4)\xi ^2{\Upsilon}' -(1\pm)\xi ^3] \\
\nonumber +(\varepsilon _{+}^2+\varepsilon _{-}^2)[(1\mp 1)\xi
^4\Upsilon +(3\pm 6) \\
 \times\xi\Upsilon -(5\mp 4)\xi ^2{\Upsilon}'-(1\mp 1)\xi
^3]\biggr]
\end{eqnarray}
for the probabilities of incoherent pair production by polarized
gamma-quanta, where
\begin {equation}%11
\Upsilon =\Upsilon (\xi)=\int _0^{\infty} \sin \biggl(\xi
t+\frac{t^3}{3}\biggr)
\end{equation}
is the upsilon function,
\begin {equation}%12
n(x)=\frac{n_0d_{pl}}{\sqrt{2\pi} u}\exp
\biggl(-\frac{x^2}{2u^2}\biggr)
\end{equation}
is the local nuclear density and
\begin {equation}%13
\frac{1}{L_{rad}}=4\alpha n_0\biggl( \frac{Z\alpha}{m}\biggr)
^2ln(183Z^{-1/3})
\end{equation}
is the radiation length. The energy dependence of optimal dichroic
polarizer length and pair production asymmetry evaluated using the
total pair production probabilities by polarized gamma-quanta
\begin {equation}%14
W_{\parallel (\perp)}=W^{coh}_{\parallel
(\perp)}+W^{inc}_{\parallel (\perp)}
\end{equation}
is given in fig. 2.

\begin{figure}[h!]% Fig.2.
\leavevmode \centering
\includegraphics[width=3.5in, height=3in, angle=0]{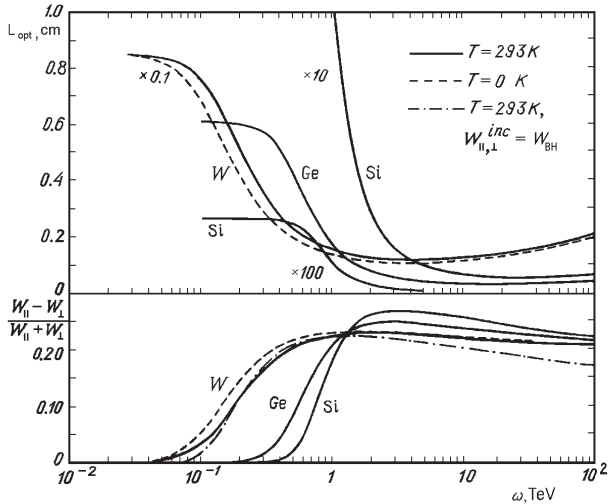}
\vspace*{-0.4cm}\caption{Energy dependence of the "optimal"
polarizator length and asymmetry coefficient of pair production by
linearly polarized gamma-quanta for different crystals and
temperatures.} \label{nurfum}
\end{figure}

A nearly quadratic proportionality of the incoherent probability
(10) to $Z$ predetermines a higher role of incoherent processes in
heavier crystals which explains the lower value of the tungsten
polarizer asymmetry. In both $Si$ and $Ge$ crystals, possessing
both lower $Z$ and much better quality, the asymmetry value
reaches its maximum at the energy close to the "threshold" energy
(5), allowing to observe the effect of uniform field dichroism in
optimal conditions as well as to use it for production and
analysis of polarized beams in the TeV energy region.

\section{Crystal birefringence}

As is well known, dichroism manifestation predetermines the
existence of birefringence. In particular, the polarization
dependence of pair production in uniform magnetic field by photons
is connected with the effect of magnetic birefringence of vacuum.
This effect is described by the non-linear QED lagrangian
\cite{heis,schw} and is considered \cite{qed2} as "one of the most
important predictions of QED which has not been verified". It also
widely manifests itself in the atmospheres of neutron stars and
white dwarfs, determining the linear polarization of x-ray
radiation emitted from their surfaces. The effect of magnetic
birefringence of vacuum causes the Cotton-Mouton effect or
transformation of the linearly polarized light into the circularly
polarized one (or vice versa).

For a long time people try to observe this effect measuring a
small ellipticity acquired by linearly polarized light in the
laboratory magnetic field. However because of the low photon
frequencies and magnetic field strengths one has to observe the
ellipticity as small as $10^{-9}$.

An alternative method to study nonlinear effect of vacuum
birefringence by observing the Cotton-Mouton effect accompanying
polarized gamma-quanta propagation through crystals was proposed
by Cabibbo \cite{cab}. The birefringence effect predicted by
Cabibbo is connected with polarization dependence of the so-called
coherent pair production and its nature considerably differs from
that of the magnetic birefringence of vacuum. This effect can be
observed at the energies considerably lower than the "threshold"
energy (5). The preliminary results \cite{apy} of its observation
are somewhat uncertain.

The LHC energy and strong fields of crystal planes provide
together really optimal possibilities to observe the vacuum
birefringence of semi-uniform averaged crystal field, the nature
of which does not principally differ from that of the magnetic
birefringence of vacuum. As is known, the relation of dichroism
and birefringence effects is expressed mathematically by the
dispersion relations. The latter allow to use the probabilities
(9) and (10) to obtain, respectively, the coherent
\begin {equation}%15
n^{coh}_{\parallel (\perp)}(\omega)=\int ^{d_{pl}}_0
n_{y(x)}(\kappa (x))\frac{dx}{d_{pl}},
\end{equation}
where
\begin{eqnarray}%16
\nonumber n_{x(y)}(\kappa)=\frac{1}{2}\Re\varepsilon _{x(y)} \\
=1-\frac{\alpha}{3\pi}\biggl(\frac{E}{\kappa E_0}\biggr) ^2\int
_0^{\omega}\frac{{\Upsilon}' (\xi)}{\xi}\biggl(\frac{\omega
^2}{\varepsilon _{+}\varepsilon _{-}}+\frac{1\mp
3}{2}\biggr)\frac{d\varepsilon _{+}}{\omega},
\end{eqnarray}
and incoherent
\begin{eqnarray}%17
\nonumber n_{\parallel (\perp)}^{inc}=n_{\omega\ll\omega_{\kappa =1}} \\
\nonumber -\frac{\sqrt{\pi}}{30L_{rad}\omega ^4}\int
_0^{d_{pl}}\frac{n(x)}{n_0}\frac{dx}{d_{pl}} \\
\nonumber \times\int _0^{\omega}d\varepsilon _{+}\biggl(\omega
^2[(1\pm 1)\xi ^4\Phi\mp 6\xi\Phi -(3\pm 4)\xi ^2{\Phi}'] \\
+(\varepsilon _{+}^2+\varepsilon _{-}^2)[(1\mp 1)\xi ^4\Phi +(3\pm
6)\xi\Phi -(5\mp 4)\xi ^2{\Phi}']\biggr)
\end{eqnarray}
contributions to the refractive indexes
\begin {equation}%18
n_{\parallel (\perp)}=n^{coh}_{\parallel
(\perp)}+n^{inc}_{\parallel (\perp)}
\end{equation}
of gamma-quanta polarized parallel and normal to a crystal plane
being nearly parallel to the gamma-quanta momentum. As is known,
the difference of refractive indexes (18) (or, more precisely, of
principal values of the refraction tensor) allows to construct a
quarter wave plate of thickness $L_{\lambda /4}$ which satisfies
the well known relation
\begin {equation}%19
\omega (n_{\parallel}-n_{\perp})L_{\lambda /4}=\frac{\pi}{2}.
\end{equation}

\begin{figure}[h!]% Fig.3.
\leavevmode \centering
\includegraphics[width=3in, height=4in, angle=0]{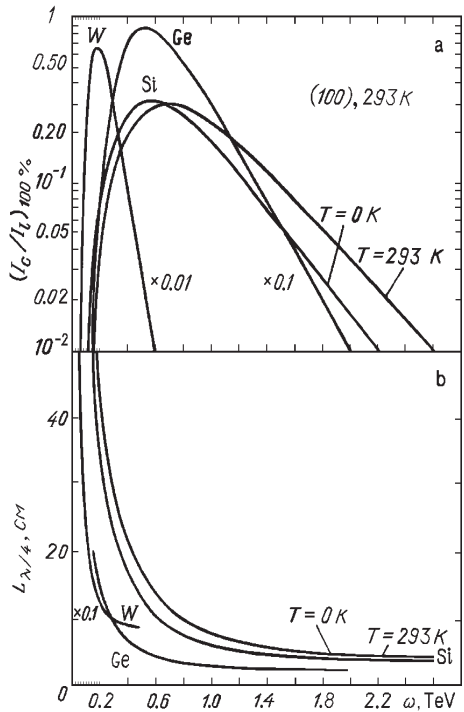}
\vspace*{-0.4cm}\caption{Energy dependence of the attenuation
coefficient and quarter wave plate thickness for different
crystals and temperatures. The dashed lines correspond to the low
energy limits \cite{bar1} of eq. (10) and (16)} \label{nurfum}
\end{figure}

The LHC energies provide really optimal conditions allowing to
observe the nonlinear vacuum birefringence effect by a complete
transformation on a linearly polarized gamma-beam to a circularly
polarized one in the quarter wave plate of modest thickness with
acceptable losses of beam intensity. The energy dependence of the
quarter wave plate thickness and the coefficient of linearly
polarized gamma-beam attenuation at this thickness is given in
fig. 3. One can easily see that the gamma-beam attenuation by a
quarter wave plate grows fast both at low and high energies due to
the increase of relative values of probabilities of, respectively,
incoherent (10) and coherent (6) pair production. Due to this
growth an optimal energy region for the observation of the
birefringence effect arises. However especially large role of
incoherent processes makes the use of tungsten quarter wave plate
not very effective even in its optimal region $\omega \sim 0.1 -
0.2 TeV$. On the other hand both $Ge$ and $Si$ quarter wave plates
possess modest lengths and beam attenuation coefficients in the
energy region $\omega \sim 0.5 - 1 TeV$, which, thus, turns out to
be the optimal region for study of the vacuum birefringence using
gamma-quantum beams which can readily be generated at LHC.

\section{Spin effects}

It should be pointed out that the effects of dichroism and
birefringence inherent to the uniform electromagnetic field can be
observed in the field of crystal planes (see fig. 1) which has
opposite directions on the opposite sides of crystal planes only
because the signs of these effects do not depend on the sign of
the field projection on the plane normal. Unfortunately, all the
effects connected with the electron spin behavior in
electromagnetic field do not possess this property and can not be
so directly observed in usual crystals. A method making it
possible to observe these effects was suggested in \cite{bar7} and
consists in using {\it bent crystals} in which channelling
particles move in the regions in which the transverse field of
particular direction dominates. Remind that bent crystals can be
also used both to clean the halo of the LHC beam and to extract it
for experiments on a fixed target.

Bent crystals also allow to observe electron self-polarization
\cite{bar8,tikh3} (see also \cite{mik1,aru}) and polarized
electron-positron pair production by gamma-quanta \cite{bar3,
bar9} (see also \cite{mik2}), positron (electron) drastic magnetic
moment decrease \cite{bar10,tikh4}, and electron spin rotation in
a circularly polarized electromagnetic wave
\cite{tikh5,tikh6,tikh7} in crystal field harmonics \cite{tikh8}.
Comparing with the crystal optical anisotropy in the gamma-region,
the theory of effects connected with the electron (positron) spin
will be additionally complicated by the necessity to consider both
the charged particle motion in crystal field and its disturbances
caused by multiple scattering and radiative cooling
\cite{tikh9,tikh10}.

Note also that another mechanism of dichroism and birefringence
connected with electron and positron scattering asymmetry by
crystal axes \cite{shu} was predicted in \cite{bar11,tikh11}.
Special attempts to study the manifestation on these effects at
LHC energies should be undertaken.

\section*{Acknowledgments}
\addcontentsline{toc}{section}{Acknowledgements}

This work is prepared in the framework of INTAS Project \#
03-52-6155.

%\newpage

\end{document}